\documentstyle[11pt,paspconf,164page,164def,psfig]{article}
\begin{document}

\title{The Caltech--Jodrell Bank VLBI Surveys}
\label{pears}

\markboth{Pearson et al.}{The CJ Surveys}

\author{T.~J. Pearson,\altaffilmark{1}
        I.~W.~A. Browne,\altaffilmark{2}
        D.~R. Henstock,\altaffilmark{2}
        A.~G. Polatidis,\altaffilmark{3}
        A.~C.~S. Readhead,\altaffilmark{1}
        G.~B. Taylor,\altaffilmark{4}
        D.~D. Thakkar,\altaffilmark{5}
        R.~C. Vermeulen,\altaffilmark{6}
        P.~N. Wilkinson,\altaffilmark{2}
    \&  W.~Xu\altaffilmark{7}}
\affil{California Institute of Technology {\em and}
       University of Manchester, Jodrell Bank}
\altaffiltext{1}{California Institute of Technology, 105-24, Pasadena, California 91125, USA}
\altaffiltext{2}{University of Manchester, Nuffield Radio Astronomy Laboratories, Jodrell Bank, Macclesfield, Cheshire SK11 9DL, UK}
\altaffiltext{3}{Onsala Space Observatory and JIVE, Chalmers Technical University, Onsala, S-43992, Sweden}
\altaffiltext{4}{National Radio Astronomy Observatory, P.O. Box O, Socorro, New Mexico 87801, USA}
\altaffiltext{5}{Oracle Corporation, Redwood Shores, California 94404, USA}
\altaffiltext{6}{Netherlands Foundation for Research in Astronomy, Dwingeloo, The Netherlands}
\altaffiltext{7}{IPAC, California Institute of Technology, 100-22, Pasadena, California 91125, USA}

\begin{abstract}
The Caltech--Jodrell Bank VLBI surveys of bright extragalactic radio
sources north of declination 35$^\circ$ were carried out between 1990
and 1995 using the Mark-II system, achieving images with a resolution
of about 1 mas at 5 GHz. The CJ1 survey (together with the older `PR'
survey) includes 200 objects with 5 GHz flux density greater than 0.7
Jy; the CJ2 survey includes 193 flat-spectrum sources with 5 GHz flux
density greater than 0.35 Jy; and we have defined a complete
flux-density limited sample, CJF, of 293 flat-spectrum sources
stronger than 0.35 Jy. We summarize the definition of the samples and
the VLBI, VLA, MERLIN, and optical observations, and present some
highlights of the astrophysical results. These include: (1)
superluminal motion and cosmology; (2) morphology and evolution of the
`compact symmetric objects' (CSOs); (3) two-sided motion in some CSOs;
(4) the angular-size--redshift diagram; (5) misalignment of
parsec-scale and kiloparsec-scale jets.
\end{abstract}

\index{Caltech--Jodrell Bank VLBI surveys}
\index{surveys}
\index{PR survey}
\index{CJ surveys}
\index{CJF survey}

\section{Introduction}

Over the last twenty-five years, VLBI imaging has made a substantial
contribution to our understanding of active galactic nuclei.  The
discovery of well-collimated, apparently superluminal jets on parsec
scales has revealed the dominant effects of relativistic beaming on
the appearance of these objects, and motivated the development of the
so-called ``unified theories'' of quasars and radio galaxies. While
detailed studies of individual objects are undoubtedly important for
understanding the origin and collimation of the jets and their
emission mechanisms, a full understanding will only come from study of
large, well-defined samples that can be subjected to statistical
analysis. This is the motivation for VLBI surveys. With surveys, we
hope to examine the full range of source morphologies and produce a
body of astrophysical data (brightness temperatures, energy densities,
flow speeds, and so forth) that can be used to develop and test
physical theories of active nuclei. 

In this paper we will present some of the results of a series of
surveys undertaken by our groups at Caltech and Jodrell Bank. We will
first summarize the sample selection and the available observational
data. We will then focus on some of the most interesting astrophysical
results.

\section{Sample Definition}

The Caltech--Jodrell Bank (CJ) surveys were conducted in three
parts. For all the surveys, sources were selected from the region of
sky with declination $>35^\circ$ and latitude $|b| > 10^\circ$. (1)
The original sample, which has now become known as the
``Pearson--Readhead'' (PR) sample, was the complete sample of 65
sources with flux density $S_{\rm 5~GHz}\ge 1.3$~Jy; 45 of these
sources have been imaged with VLBI at 5 and 1.6 GHz (Pearson \&
Readhead 1981, 1988; Polatidis \etal\ 1995). (2) The CJ1 survey
extended the PR sample down to $S_{\rm 5~GHz}\ge 0.7$~Jy, adding 135
sources for a total of 200. Eighty-two of the additional sources have
been imaged with VLBI at 5 and 1.6 GHz (Polatidis \etal\ 1995; Thakkar
\etal\ 1995; Xu \etal\ 1995). (3) For the CJ2 sample, we further
decreased the flux-density limit to $S_{\rm 5~GHz}\ge 0.35$~Jy, but
with the restriction that sources should have a flat spectrum ($\alpha
> -0.5$), which eliminates extended sources that are difficult to
image with VLBI. The CJ2 sample contains 193 sources, all of which
have been imaged with VLBI at 5~GHz (Taylor \etal\ 1994; Henstock
\etal\ 1995). 

We have more recently defined a complete flux-density limited sample
of flat-spectrum sources stronger than 0.35 Jy, which we call the
Caltech--Jodrell Bank flat-spectrum sample (CJF). This includes 293
extragalactic sources. For this sample, we have re-examined the
single-dish surveys from which the sources were selected, and included
18 sources that were not included in PR, CJ1, or CJ2 (Taylor \etal\
1996). The new selection is based on the 1.4 and 4.85 GHz Green Bank
surveys. Thus we have complete samples of {\it all} sources down to
0.7~Jy (PR+CJ1), and of {\it flat-spectrum} sources down to 0.35~Jy
(CJF).

\section{Observations}

The primary goal of the surveys was to image the parsec-scale
structure in the objects at 5 GHz. The original observations were VLBI
snapshots with a global array, using the old Mark-II narrow-band
recording system. We were pleased to find that reliable images of
compact objects could be made with only three 20-minute scans spaced
in hour angle.  In
addition, we made 1.6-GHz VLBI images of the sources in the PR and CJ1
samples.  All the images are presented in the papers cited in the
previous section.

To compare the parsec-scale structure with the larger-scale structure
in these objects, we have made (or obtained from the literature)
images with the VLA and MERLIN at 5 and 1.6 GHz. We have also
assembled radio spectra of the sources from the literature (Herbig \&
Readhead 1992; Xu \etal\ 1995; Henstock \etal\ 1995; Taylor \etal\
1996).

We are making extensive optical observations to obtain redshifts for
all the sources, using the Palomar 200-inch, Isaac Newton, William
Herschel, and Keck telescopes (Henstock, Browne, \& Wilkinson 1994; Xu
\etal\ 1994; Lawrence \etal\ 1996; Vermeulen \& Taylor 1995; Vermeulen
\etal\ 1996; Henstock \etal\ 1997); for the PR sample, we have
obtained high-quality, calibrated spectra which will allow statistical
investigations of the relationship between the optical spectrum and
the parsec-scale structure (Lawrence \etal\ 1996). We are approaching
completeness: CJF is currently 97\% optically identified, and we have
redshifts for 92\% of the objects.  The redshifts range up to 3.886.

An important part of the project is continued VLBI monitoring to
measure internal proper motions in the parsec-scale radio
sources. Almost all of the CJF sources have been observed at at least
two epochs, and we are part-way through observations for a third
epoch, with a typical separation between observations of about two
years. For most of the PR sources, we have four or more observations
spread over nearly 18 years. All the recent observations have been
made with the VLBA in snapshot mode.

\section{Highlights}

The CJ surveys have produced a large body of observational data that
we hope will be of value for a wide variety of astrophysical
investigations. In this paper, we have space only to mention some of
the investigations that we have undertaken ourselves, and give
references to more extensive presentations.

\subsection{Superluminal Motion}\index{superluminal motion}

The great majority of sources imaged in the CJ surveys show a
parsec-scale jet on one side of a flat-spectrum, compact core. In the
standard model, the asymmetry is attributed to relativistic beaming
and the core is close to, though probably not coincident with, the
center of activity.  Several of the sources in the CJ surveys, and
especially those in the PR sub-sample, have been monitored extensively
to follow the evolution of brightness peaks (usually called
``components'') as they move out along the jet.  This is not the place
to discuss these detailed studies of individual objects, but some
results are apparent: components show a wide range of apparent
velocities; different components in the same source can show different
velocities; components can accelerate, decelerate, merge or split; and
in some cases a stationary component can coexist with moving
components. It is quite clear that the components are not discrete
objects following ballistic paths; the apparent motions that we
measure are probably phase velocities of a moving disturbance and may
not be simply related to the velocity of the underlying jet flow.

A major goal of the CJ surveys is to make a statistical study of the
apparent speeds, in order to understand the general properties of the
phenomenon rather than the details of an individual object. We may
also be able to place constraints on $q_0$ and $H_0$ (Vermeulen \&
Cohen 1994; Lister \& Marscher 1997)\index{Hubble constant}.  Analysis
of the second and third epochs is not yet complete: what follows is an
update on the preliminary results presented by Vermeulen (1995, 1996),
based on 81 objects. The distribution of apparent velocities
($\beta_{\rm app}$), which depends on $h = H_0/100$
km\,s$^{-1}$\,Mpc$^{-1}$ and $q_0$, shows an abundance of velocities
in the range from zero to $5h^{-1}c$, with only a tail of the
distribution reaching up to $\sim 10h^{-1}c$. A substantial fraction
($\sim 25\%$) of the objects show stationary features or have only
upper limits on velocity. In a beamed, randomly oriented sample of
jets with speed $\beta c$ (Lorentz factor $\gamma$), most sources
should show $\beta_{\rm app} \approx \beta\gamma$, so this result
shows that either the pattern motions we measure are different from
the underlying bulk flow (by a factor up to 10), or there is a range
of more than a decade in jet Lorentz factors over the sample, or
both. The preponderance of $\beta_{\rm app} > 5h^{-1}$ in the earlier
literature probably reflects a bias toward the faster sources, which
are easier to measure. There are no significant differences between
the average apparent speeds for galaxies ($\beta_{\rm app} = 2.1$), BL
Lac objects ($\beta_{\rm app} = 2.3$), and quasars ($\beta_{\rm app} =
3.2$); here we have ignored upper limits.  A Kolmogorov-Smirnov test
gives a probability of 30\% that the $\beta_{\rm app}$ values for
quasars (44 objects) and BL Lacs (8 objects) are taken from the same
distribution, in contrast to the result of Gabuzda \etal\ (1994).
\index{BL Lac objects}

There is a clear and very intriguing correlation between apparent
velocity $\beta_{\rm app}$ and luminosity at 5~GHz (observed
frequency), calculated assuming isotropic emission. While low
$\beta_{\rm app}$ can be found at any observed luminosity, the largest
$\beta_{\rm app}$ occur only at the highest luminosity (Vermeulen
1996). Recent simulations by Lister \& Marscher (1997) show that this
effect can be obtained as a sort of Malmquist bias from the interplay
between Lorentz factor distributions weighted to low values and the
shape of the unbeamed luminosity function and its cosmic evolution. It
is also possible that the Lorentz factor may be correlated with
unbeamed luminosity.

\subsection{Compact Symmetric Objects}\index{compact symmetric objects}

Between 5 and 10\% of the sources in the CJ samples are {\it compact
symmetric objects} (CSOs), in which high-luminosity radio emission
regions are seen on {\it both} sides of the center of activity on
scales less than one kiloparsec (Wilkinson \etal\ 1994). This class
includes the {\it compact double} sources earlier recognized by
Phillips \& Mutel (1982). A serious impediment to our understanding of
these objects has been the difficulty of determining the exact
location of the central engine---the best candidate is a compact,
flat-spectrum component, but it is not always the brightest feature
near the middle of the structure (Taylor, Readhead, \& Pearson 1996).
Although the overall structure is frequently symmetric about this
point, there may be bright asymmetric emission perhaps associated with
a jet, as in 0710+439\index{0710+439} and
2352+495\index{2352+495}. The correct classification of some objects
remains in doubt because of the difficulty of pinpointing the center
of activity (e.g., CTD\,93, p.~\ref{shaffe}). In general, the CSOs
have no larger-scale extended structure, have low polarization and low
variability, and are identified with galaxies rather than quasars. The
advance speed of the outer ``hot spots'' is low ($< c$).

\index{evolution of radio sources} \index{confinement} The objects
appear to be small versions of the extended double radio galaxies, and
the question remains whether they are young precursors of
kiloparsec-scale objects, are short-lived objects that will never
achieve kiloparsec size, or are older objects that are unable to
expand owing to strong confinement by the interstellar medium. If the
hot spots are confined by ram pressure, we can use the observed
pressure in the hot spots to constrain the advance speed $v_{\rm a}$
and the density $\rho_{\rm ext}$ of the confining medium (Readhead
\etal\ 1996a). For example, in 2352+495 we find $\rho_{\rm ext} =
10(v_{\rm a}/0.02c)^{-2}$ cm$^{-3}$, and hence the age of the object
is $\sim 3000 \rho_{\rm ext}^{0.5}$ yr. A variety of observations
place limits on the density of \Hii\ and \Hi\ in the nuclear regions
of the galaxy, $\rho_{\rm ext} < 10^3$ cm$^{-3}$, which imply an age
$< 10^5$ yr. A larger density and age are possible if the medium is
molecular (H$_2$), but an age as large as $10^6$ yr would require an
implausibly large mass of gas within the central 200 pc. We conclude
that the CSOs are young objects, typically of age $10^4$~yr at a size
of 50~pc, with $v_{\rm a} \approx 0.02c$.  It is possible that the
CSOs evolve into CSS doubles (MSOs), and then into large-scale FR II
objects (Hodges \& Mutel 1987; Fanti \etal\ 1995; Readhead \etal\
1996b; De Young, these Proceedings, p.~\ref{deyou}; O'Dea, these
Proceedings, p.~\ref{odea}).  The numbers of objects show that, if
the expansion speed remains constant, the luminosity must decrease
roughly as $R^{-0.3}$ and the external density must decrease as
$R^{-1.3}$, where $R$ is the distance of the hot spot from the center
of activity. On this model, the CSOs that we know of are the
precursors of lower-luminosity FR-II objects, not the rare, highly
luminous ones.

\begin{figure}[t]
\cl{\psfig{figure=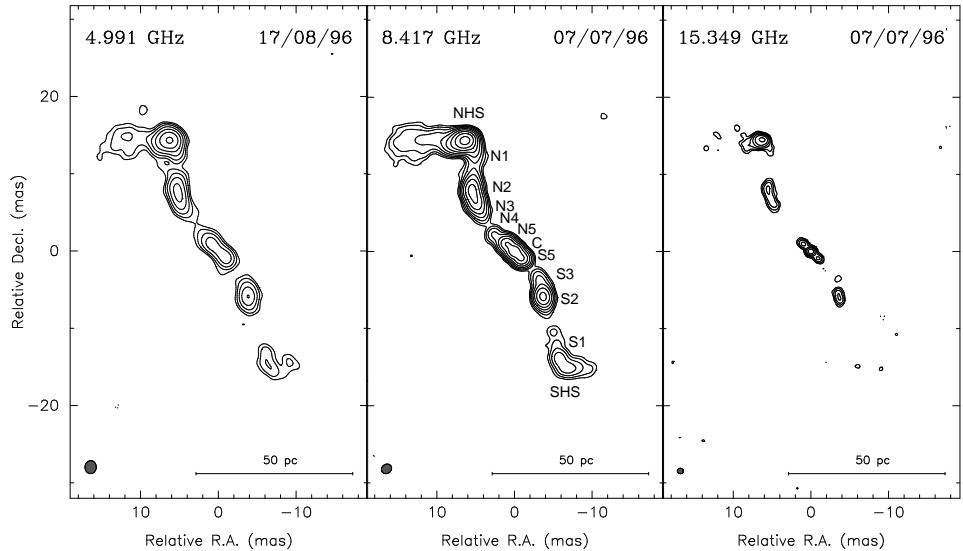,width=\textwidth}}
\caption{ VLBA images of 1946+708. The core component is labeled
C. Outward motions have been measured for the pairs of components
N2/S2 and N5/S5 (from Taylor \& Vermeulen 1997).}
\label{pearsf1}
\end{figure}

\subsection{Bidirectional Motion}\index{jets!bidirectional motion}

A particularly interesting result from the CJ surveys is the discovery
of bi\-directional relativistic jets in the radio galaxy 1946+708
(Taylor, Vermeulen, \& Pearson 1995; Taylor \& Vermeulen
1997).\index{1946+708} This is a CSO at a redshift $z=0.101$.  Several
components can be identified, symmetrically placed on opposite sides
of a compact, inverted spectrum component that is presumably
the core (Figure~\ref{pearsf1}). The two outer ``hot spots'' show an
insignificant separation rate of $0.22 \pm 0.3$ mas~yr$^{-1}$,
consistent with the low advance speed seen in other CSOs. If the
images are aligned using the hot spots, intermediate components in the
jets show significant motion away from the core component, which is a
good indication that this component represents the true center of
activity.  When motion is detected on only one side of the core, one
can measure one parameter involving the intrinsic speed $\beta$ and
the angle to the line of sight $\theta$: $\beta_{\rm app} =
\beta\sin\theta/(1 - \beta\cos\theta)$. With components on both sides,
however, one can estimate both $\beta$ and $\theta$. If the two
components were emitted at the same time with the same speed, the
ratio of their distances from the core (arm-length ratio) should be
$(1 +\beta\cos\theta)/(1-\beta\cos\theta)$, and the apparent speed of
separation of the two components should be $2\beta c \sin\theta /(1 -
\beta^2\cos^2\theta)$. Conversion of angular separation rate to linear
separation speed involves the distance of the source and is thus
dependent on $H_0$. In this particular object, the measurements are
only consistent with the simple model if $65^\circ < \theta <
80^\circ$ (i.e., the axis of the source is at a substantial angle to
the line of sight), and $\beta \approx 0.6$ for reasonable values of
$H_0$. If the jets are relativistic ($\beta \approx 1$), then $H_0 =
37 \pm 6$ km\,s$^{-1}$\,Mpc$^{-1}$.\index{Hubble constant} The
prospects are encouraging that future observations of bidirectional
jets in this and other objects can be used to test the relativistic
jet model in some detail, and possibly place useful constraints on
$H_0$ and $q_0$. A second such object, NGC\,3894 (1146+596) has
recently been found in the CJ sample (Taylor, Wrobel, \& Vermeulen
1997).\index{NGC\,3894}\index{1146+596}

\subsection{Angular-Size -- Redshift Test}\index{angular-size--redshift test}\index{cosmology}

The CJ surveys provide excellent, homogeneous samples for studying
cosmology by way of the angular-size--redshift test for compact radio
sources. The possibility of measuring $q_0$ in this way has attracted
considerable attention since Kellermann (1993) first suggested it and
showed that the data were consistent with $q_0 = 0.5$. A similar
result was obtained by Gurvits (1994) using angular sizes estimated
from visibility data rather than measured from images. The main
drawback of the method is that it is not clear that the measured
angular size corresponds to a ``standard rod'' whose effective linear
size is independent of redshift, although Kellermann's results suggest
empirically that it does. Several authors (e.g., Dabrowski, Lasenby,
\& Saunders 1995) have shown that Kellermann's result, based on 82
sources, is not statistically compelling. We have therefore repeated
Kellermann's analysis with the complete CJF sample (Wilkinson \etal\ 1997).

Following Kellermann, we have measured the angular size between the
peak in the image (the core) and the most distant component (peak)
exceeding 1\% of the core brightness (Kellermann used 2\%). Unlike
Kellermann, we have convolved each map to the same {\it linear}
resolution corresponding to an angular resolution of 1.5 milliarcsec
at redshift $z=1.25$, assuming a value of $q_0$. This should reduce
the bias due to the dependence of angular resolution on redshift. At
any redshift there is a large scatter in the angular size
measurements, but we find that the median angular size decreases as
$z$ increases up to $\sim 0.5$, and flattens off at higher $z$, in
qualitative agreement with the results of Kellermann and Gurvits. The
upper envelope of the distribution does not decrease with redshift.
However, as emphasized by Wilkinson \etal, there are several possible
effects that must be understood before the data can be used to obtain
a reliable estimate of $q_0$, including the dependence of measured
angular size on luminosity and the possible effects of evolution.

\subsection{Misalignment-Angle Distribution}\index{jets:bent}\index{misalignment}

A surprising result that emerged from the PR survey was the discovery
of a bimodality in the distribution of position-angle differences,
$\Delta$PA, between the nuclear (parsec-scale) radio structure and the
extended (kiloparsec-scale) structure. The distribution of $\Delta$PA
shows two distinct peaks, near $0^\circ$ (aligned objects) and near
$90^\circ$ (misaligned objects). This result was later confirmed in
larger samples (Wehrle \etal\ 1992; Conway \& Murphy 1993; Appl, Sol,
\& Vicente 1996), and the peak near $90^\circ$ was shown to be due
primarily to objects of high optical polarization (Impey \etal\ 1991).
With the completion of the CJ surveys, we now have larger, homogeneous
samples, with which to investigate this phenomenon further. We have
compiled the $\Delta$PA distribution for the 87 objects from PR and
CJ1 for which we have adequate VLBI and VLA images, and we have made
new measurements of the optical polarization of the CJ1 sources (Xu
\etal\ 1997); unfortunately we have not yet made a similar analysis of
CJ2. We have confirmed the bimodality of the distribution, and find
that large misalignment angle is associated with high optical
polarization, prominent parsec-scale core component, and flat
spectrum---``blazar'' properties that are usually attributed to
orientation of the jet toward the observer and relativistic beaming.
The bends that produce the $90^\circ$ peak in the distribution occur
over a wide range of scales in different sources, from a few parsecs
to a few kiloparsecs; in a few cases, a jet can be traced continuously
through the bend from parsec to kiloparsec scales.

It is now clear that the bimodality is real, and is connected in some
way with relativistic motion of the nuclear emission regions.
Successful models of active galactic nuclei must account for this
phenomenon. Possible explanations include helical jets with
relativistic beaming (Conway \& Murphy 1993), ram pressure bending
(Pearson \& Readhead 1988), and warped accretion disks (Appl \etal\
1996). We prefer the geometrical model of Conway \& Murphy, in which
the visible part of the nuclear jet lies on the surface of a cone and
relativistic beaming favors lines of sight close to the cone surface
where $\Delta$PA $\sim 90^\circ$ will be observed. Various physical
conditions might give rise to such a geometry: a helical jet driven by
a binary black hole is an interesting possibility, but it is also
possible that the visible emission is due to a helical instability on
the surface of a wider conical jet.\index{jets!helical}

\section{Conclusion}

The Caltech--Jodrell Bank surveys have yielded a large body of
observational data that should be of use both for identifying
individual sources of particular interest and for a wide variety of
astrophysical investigations. Some of the results could have been
predicted, for example the high incidence of superluminal motion, but
the study of complete samples is yielding new data that will allow us
to improve our understanding of this phenomenon. The identification of
a large number of CSOs has shown that these objects really are a
distinct population, not dominated by relativistic beaming, and it is
clear that detailed studies of these objects will cast light on the
early stages of radio source evolution. The prospects for cosmological
studies are good, both for the direct measurement of $H_0$ with
bidirectional jets and for the determination of $q_0$ through
statistics of superluminal motion and angular sizes, although, as in
all such studies, the effects of evolution and of observational bias
need to be thoroughly understood. There are plenty of directions for
future work, both in extending surveys to fainter sources and other
areas of the sky, and in following up the results obtained so
far. Indeed, several large-scale systematic surveys are now under way
with the EVN, the VLBA, and HALCA, and many of them are described in
other papers presented at this meeting.

\footnotesize\acknowledgments 

The CJ surveys represent one of the largest
VLBI projects carried out prior to the completion of the VLBA, and we
are very grateful to the personnel of the participating observatories
and of the Caltech/JPL correlator for making the surveys possible.
The work at Caltech was supported by the National Science Foundation
(AST-9117100 and AST-9420018).
\NRAOcredit

\end{document}